# Design an Ontology for Cognitive Business Strategy Based on Customer Satisfaction

**Neda Bagherzadeh**
[Department of Computer Engineering
North Tehran Branch, Islamic Azad University, Tehran, Iran
ORCID: 0000-0001-8533-2386, n.bagherzadeh.r@gmail.com]

**Saeed Setayeshi**
[Department of Physics and Energy Engineering
Amirkabir University of Technology, Tehran, Iran
ORCID: 0000-0002-0569-6946, setayesh@aut.ac.ir]

**Samaneh Yazdani**
[Department of Computer Engineering
North Tehran Branch, Islamic Azad University, Tehran, Iran
ORCID: 0000-0001-7806-5152, s.yazdani87@gmail.com]

**Abstract:** Ontology is a general term used by researchers who want to share information in a specific domain. One of the hallmarks of the greatest success of a powerful manager of an organization is his ability to interpret unplanned and unrelated events. Tools to solve this problem are vital to business growth.

Modern technology allows customers to be more informed and influential in their roles as patrons and critics. This can make or break a business. Research shows that businesses that employ a customer-first strategy and prioritize their customers can generate more revenue. Even though there are many different Ontologies offered to businesses, none of it is built from a cognitive perspective.

The objective of this study is to address the concept of strategic business plans with a cognitive ontology approach as a basis for a new management tool. This research proposes to design a cognitive ontology model that links customer measurement with traditional business models, define relationships between components and verify the accuracy of the added financial value.

**Key Words:** Business model Ontology, Cognitive strategy, Cognitive ontology, Customer satisfaction, BEL model
**Category:**
**DOI:**

## 1 Introduction

Scientific research on business models is essential for several reasons. Firstly, business model concepts are widely discussed, and a common understanding of what defines this concept is necessary. Secondly, it can be an appropriate and fundamental method for management and engineering tools to respond to a growing and dynamic business environment.

Recent research has shown that a company needs to use customer-centric culture to succeed [Cronin 2001]; [Hennig-Thurau 2004]. A customer-oriented approach is a milestone in the development of the theory and practice of marketing. The importance of customers has led to more effort for customer retention and the use of customer-oriented metrics such as satisfaction and loyalty [Roy 2015]. Loyal customers are reported to devote a higher share of their wallet to the organization and are more likely to introduce the organization to others [Zeithaml 2000].

New technologies, such as the Internet, have simplified information access; hence, using these technologies, customers became more aware of purchasers than before. Besides, digital technologies have facilitated communication between customers, so people can share their experiences of using products and services with others [Urban 2005].

According to the expectation disconfirmation theory proposed by Oliver, customer satisfaction stems from a comparison between perceived performance and expectations [Oliver 2014]. Satisfying a customer is likely to result in customer loyalty [Fornell, Johnson et al. 1996]. If a company adopts a customer advocacy strategy as its main business approach, providing honest and complete information about products and services, there is a higher chance of earning trust and achieving higher profit and growth [Reichheld 1993, Urban 2005, Roy 2015].

Given that most research on business plans is conducted at a non-conceptual and sometimes ambiguous level, and considering that cognitive sciences have not been sufficiently focused on in this field. Absolutely improving business operations through the use of cognitive sciences can be highly effective. Cognitive sciences can help us understand how people think, learn, and make decisions, which can be leveraged to enhance various aspects of business, such as decision-making processes, customer interactions, and employee training. By integrating cognitive science insights, businesses can create more intuitive systems, better predict consumer behaviour, and optimize overall performance.

The aim of this paper is to define a general cognitive model to describe business models and evaluate it [Bagherzadeh, Setayeshi et al. 2023]. The proposed model is evaluated using the BEL model, and the results are analysed and reviewed. The cognitive ontology of the business model presented in this paper, and the tools developed based on it, are the first steps toward facilitating management in uncertainty.

This paper is organized as follows: first, the core concepts used in the study are reviewed. In this section, customer metrics, cognitive ontology, the business model ontology used in this research and BEL model are described. Second, an ontology for cognitive business strategy based on customer satisfaction is proposed. This section contains the design of the business model ontology. Next, evaluating the new ontology using the BEL model, and results analysis are proposed. Finally, the conclusion and directions for future research are presented.

**1.1 Ontology**

Ontology, in the realm of philosophy, pertains to the study of the nature of existence and being. Within the field of computer science, ontology serves as a framework for the sharing and reuse of information resources between software systems and human users. It encompasses the definition of terms as well as the semantic relationships among them in a manner that is machine-readable and interpretable.

An ontology constitutes an explicit and formal representation of the concepts and relationships within a specific domain. A primary objective in the creation and development of ontologies is to establish a common understanding of the informational structure shared by both human and machine agents. Additionally, ontologies facilitate the reuse of knowledge; a general ontology can be extended and adapted to generate more specific ontologies tailored to particular fields of study.

**1.2 Cognitive science**

Cognitive science is an interdisciplinary scientific field that studies the nature of mental activities such as thinking, classification, and the processes that enable these activities. It incorporates elements from psychology, linguistics, philosophy, neuroscience, computer science, and anthropology. The main goal of cognitive science is to understand and formulate the principles of intelligence, with the hope of enhancing our comprehension of the mind and learning processes.

In the context of artificial intelligence, cognitive science aims to leverage the computational power of computers to understand human thinking. Emotion-aware applications must be designed to be flexible, accommodating a wide range of users. Personalization is essential for more effective interaction, better regulation, and the acceptance of developed systems.

**1.3 Cognitive ontology**

Ontological background knowledge is integral to many modern information technology applications. The vast quantities of data, information, sentiments, and opinions available on the web necessitate systematic sorting and structuring to facilitate efficient information retrieval and acquisition. Ontology-based frameworks can integrate diverse knowledge sources, effectively bridging the gap between human and machine understanding of emotions [Brenga, Celotto et al. 2015].

Given the significance of this issue, there is a need for a standardized method to express emotions that can be seamlessly integrated, shared, and utilized to enhance user experiences in various applications. Due to the complexity of these data, employing semantic models based on ontology is a logical approach.

Numerous models for understanding feelings and emotions have been proposed in the literature, most of which describe emotion maps with potential connections between various terms. By leveraging cognitive ontology, it is possible to establish coordination and integration among these concepts, fostering a more unified understanding of emotional data.

**1.4 Business plan**

A business plan is a comprehensive and descriptive document outlining the business activities of an institution, group, or entrepreneur. Entrepreneurship and the initiation of any new business venture often entail significant risks. Thus, the necessity of employing an appropriate plan and model to conduct thorough investigations and analyses prior to commencing operations, in order to minimize these risks and potential damages, is indisputable.

Broadly, a business model can be categorized into four main components:
- Product Innovation: This block articulates the value proposition of a company.
- Customer Relationship: This block explains how a company interacts with its customers and the type of relationships it aims to establish.
- Infrastructure Management: This block details the essential activities, resources, and partners required to support the first two blocks.

- Financial Aspects: This block describes the revenue streams and pricing mechanisms of a company, essentially explaining how the company will generate income through the other three blocks.

The ontology of the business model, with these divisions, is depicted in [Fig. 1] [Osterwalder and Pigneur 2002].

The integration of cognitive concepts can significantly enhance a business plan. Evaluating and reviewing this planning from a cognitive perspective can lead to more precise outcomes and improved performance. However, there remains an absence of adequate cognitive management tools for understanding, mapping, and analysing business logic.

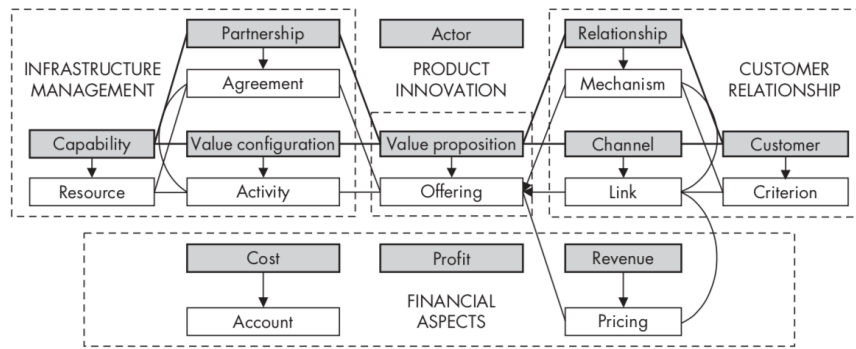

*Figure 1: The Ontology of the Business Model and Its Four Buildings*

### 1.5 Cognitive Business Plan Ontology

Customer satisfaction provides crucial criteria for business owners and marketers to manage and improve their enterprises. Customer satisfaction is a cognitive process that necessitates a cognitive strategy. Research has demonstrated that cognitive strategy is synonymous with strategic cognition.

Therefore, the decision-making process must be modelled dynamically to effectively analyse human behaviour in real-world scenarios. This study introduces a model wherein customer preferences are influenced by various factors such as thoughts, perceptions, values, emotions, information, and environmental conditions. Considering these factors, a business model can be planned with greater accuracy and effectiveness. Customer satisfaction naturally leads to increased loyalty and system efficiency. The business model ontology serves as a conceptual tool that explains the business logic of a company. Ideally, this model can form the foundation for new management tools in strategy and information systems, facilitating the development of a comprehensive cognitive business strategy [Bagherzadeh, Setayeshi et al. 2023].

### 1.6 BEL Model

BEL model is a computational model inspired by the emotional learning processes in the human brain, particularly the limbic system and amygdala [Morén and Balkenius 2000]. This model inspired by the structure of the brain in handling input data and the involvement of emotions. The limbic system is the primary part of the brain responsible for detecting emotions. BEL model is one of the most powerful models of emotional learning and includes four main subsystems: the amygdala, orbitofrontal cortex, thalamus, and sensory cortex. These subsystems form the basis of the used model and have a simple structure. Figure 2 provides an overview of the BEL model utilized in this article [Morén and Balkenius 2000].

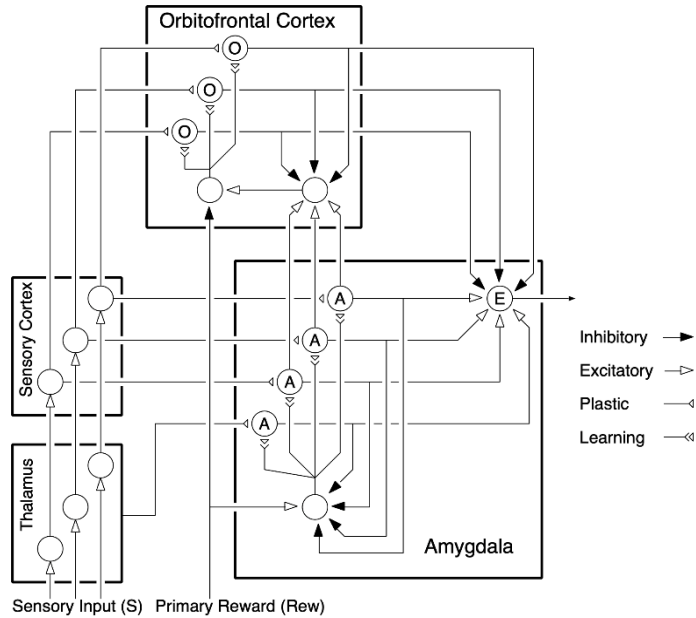

*Figure 2: An overview of the BEL model*

At the top of this model, the orbitofrontal region is located, while the amygdala region is positioned at the bottom right. On the left side, the thalamus and sensory-cortex modules can be found. In this version of the model, the thalamus and sensory-cortex parts serve primarily as placeholders. Sensory inputs (S) enter through the thalamus module, where the input to the amygdala is computed as the maximum value among all inputs. Additionally, a primary reward signal (Rew) is directed to both the amygdala and orbitofrontal regions.

This system operates at two levels: the amygdaloid part and the orbitofrontal part. The amygdaloid part learns to predict and respond to a given reinforcer. Once a connection is learned, it is permanent, allowing the system to retain emotional connections indefinitely. The orbitofrontal system, on the other hand, monitors discrepancies between the base system's predictions and the actual received reinforcer, learning to inhibit the system's output proportionally to the mismatch.

These subsystems receive partially different inputs. The base system receives finely discriminated inputs from the sensory cortex and a coarse signal (Th) from the thalamus. The sensory cortex also receives its inputs from the thalamus and is responsible for subdividing and discriminating the coarse input from the thalamus. The thalamic input, a low-level stimulus signal, is present even in the absence of higher cortical areas. Although this input may appear unnecessary or even harmful, evidence suggests that this pathway exists.

The BEL model is one of the practical and successful artificial intelligence models that has been utilized in numerous articles. For instance, in source [Fakhrmoosavy, Setayeshi et al. 2018], this method has been employed for earthquake magnitude and fear prediction. Sources [Farhoudi, Setayeshi et al. 2017] and [Motamed, Setayeshi et al. 2017] have utilized this method for speech emotion recognition. Source [Parsapoor, Bilstrup et al. 2014] has applied this model for a data-driven approach to predict geomagnetic storms.

## 2 Literature review

Nowadays, sentiment analysis has gained increased attention in the scientific community. Existing approaches to multidimensional sentiment analysis primarily focus on mapping multi-sentence information from text sections where opinions are explicitly described, such as phrases, words, and their overlapping frequencies.

Numerous works have been conducted in the field of ontology. Previous research, such as [Yaakub, Li et al. 2013], has presented a feature ontology that utilizes a multi-dimensional model to integrate customer features and their opinions on products. In the field of ontology, resources such as the foundational ontology [Masolo, Borgo et al. 2003] DOLCE have been widely developed for the analysis of social entities. Additionally, other works have addressed areas such as formal ontology of collections, formal ontology of functions, and formal ontology of processes [Turki, Kassel et al. 2016].

Previous research has focused on sentiment analysis at various levels, but domain knowledge, contextual relationships, and multilingual applications have not been considered together in these studies [Tao and Liu 2017]. In source [Dragoni, Poria et al. 2018], OntoSenticNet has been proposed as a conceptual model that supports the structural analysis of sentiments from multi-purpose sources based on SenticNet, a general knowledge base for sentiment analysis. Another model of Emotion Ontology, named [MFOEM], has been created to support a structural representation of mental functions, including cognitive processes and traits such as intelligence [Dragoni, Poria et al. 2018].

In the literature, several models for feelings and emotions exist; most describe emotions with conceptual maps that illustrate any possible connections between terms. WordNet-Affect is an emotion lexicon based on WordNet [Strapparava and Valitutti 2004].

Despite the significant advancements in emotion detection, the field of emotion ontology still requires further development. Additionally, extensive research has been conducted by scholars in quality management to enhance various business processes, thereby improving their outcomes and performance with a focus on customer needs. In strategic management, source [Lorino 2003] aims to identify strategic resources and processes to enhance organizational performance. Other studies have also been carried out in the fields of information systems [Melão and Pidd 2000] [Nurcan, Etien et al. 2005], workflow management [Coalition 1999], business process management [Hammer and Champy 2009], and business innovation [Davenport 1993] to improve and optimize internal business processes as well as processes shared with other organizations.

Numerous articles have been presented concerning the components of business models, and several business model ontologies have been proposed, such as the STOF model [Bouwman, Faber et al. 2008], the VISOR model [El Sawy and Pereira 2013] and the Business Model Canvas [Osterwalder and Pigneur 2002]. In source [Andersson, Bergholtz et al. 2006], a reference ontology for business models is proposed using concepts from foundational business model ontologies such as the Business Model Ontology (BMO), Resources Events Agents (REA), and e3-value. Source [Hamrouni 2021] applied Case-Based Reasoning (CBR) as a problem-solving paradigm that uses knowledge of relevant past experiences (cases) to interpret or solve new problems for software tools of business model development to supporting business model design, sustainability, and innovation.

The core concepts in the reference ontology pertain to role players, resources, and the transfer of resources between these entities. Source [Mohd Nasir 2024] advocates the adoption of the Unified Ontology Approach (UOA) as a comprehensive and flexible framework for Business Model Ontology (BMO) development. Source [Upward and Jones 2016] presents a sustainable ontology for business models by defining an organizational framework compatible with natural and social sciences, addressing the weaknesses of previous models and providing the necessary relationships and efficiency.

Another source [López and La Paz 2017] offers an ontology for strategic information systems planning, which allows adaptability by adding categories to classifications or reducing classifications within the ontology.

Given the aforementioned points, the business plan has been extensively reviewed and developed from various aspects and has improved in recent years. Some of the positive points addressed in various sources include the following:

- Managers involved in business model innovation have faced a series of cognitive challenges. A cognitive-oriented perspective has been utilized for various analyses of visual experience, examining how visual tools can support business model innovation at the cognitive level.
- By establishing a connection between process models and ontology through the standardization of terms, business models have been constructed automatically with ontological concepts during the design phase.
- Adaptability has been provided for some ontologies.
- The cognitive approach to managing business models and their underlying logic has been examined using personal construct theory and the common network method, which is a novel approach to discovering central factors in entrepreneurial business model management.

The creation and expansion of cognitive ontologies for business models are undoubtedly among the most effective steps toward improving the quality of business models. Although business models have been cognitively reviewed, a cognitive ontology for these models has not yet been introduced.

Therefore, in this research, the ontology of the business plan has been examined from a cognitive perspective. In this paper, the cognitive ontology for the business plan presented in source [Bagherzadeh, Setayeshi et al. 2023] is tested using the BEL model, and the results obtained are analysed.

## 3 Customer Satisfaction and Cognitive Processes in Business Models

Customer satisfaction provides business owners and marketers with criteria to manage and improve their enterprises. Thus, it can be summarized that customer satisfaction is a cognitive process, and a cognitive strategy must be defined for it. In previous research, scientists have equated cognitive strategy with strategic cognition.

The actual conditions of customer decision-making differ significantly from those depicted for rational humans. The motivations or goals of customers are not limited to personal gain but also depends on altruism, loyalty, hostility, and spite; which conventional management rules like cost-benefit analysis do not apply to.

Conditions and environments are neither fixed nor unchanging. Both agents influence environmental conditions, and conditions affect agents and their behaviour. Therefore, the decision-making process must be dynamically modelled to accurately explain and analyse human behaviour and realities. This study introduces such a model. In this approach, customer preferences are influenced by various factors, such as thoughts and perceptions, values and emotions, information and knowledge, and even environment and conditions. Information and knowledge not only affect preferences but also have a reciprocal relationship with thoughts and perceptions. Information and knowledge influence preferences and are influenced by them. Values and emotions have various dimensions and are not the same among different individuals, changing through learning and experience, and have a two-way relationship with thoughts and perceptions. Environment and conditions also directly and indirectly (through influence on preferences) affect decision-making.

By considering such factors related to customer behaviour and decision-making, and addressing customer needs from a cognitive perspective, a business model can be more precisely and effectively designed. Naturally, customer satisfaction leads to increased loyalty, resulting in higher profits and improved system efficiency. [Fig. 2] symbolically shows this cognitive process (factors influencing customer decision-making).

## 4 A Business Model Ontology of a Cognitive Strategy Based on Customer Satisfaction

The purpose of this study is to demonstrate the importance of customers for any business to succeed. Initially, the conventional business model ontology is considered the main framework of this research; furthermore, the customer metrics, presented in Section 2.1, are combined with the business model to examine the impact of customers in a business. Regarding the conventional business model ontology, the CUSTOMER INTERFACE pillar is the customer-related pillar in this ontology, and consequently, to provide a new business model ontology that holds the customer-related parameters, this pillar should be analysed. The Relationship block, which includes a set of mechanisms, determines the relation a company wants to establish with its customers. Assuming that a company adopts the customer importance strategy as its business strategy, it builds its relationship with its customers based on this mechanism.

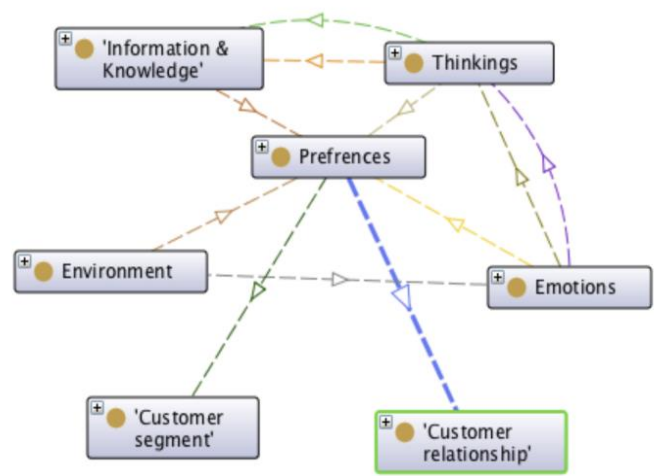

*Figure 3: Ontology of Cognitive Processes Effective in Customer*

The new business model ontology as described in the previous article [Bagherzadeh, Setayeshi et al. 2023] is composed of the nine building blocks of Osterwalder's ontology [Osterwalder, 2004, p. 80] and CUSTOMER INTERFACE part with 5 blocks. Hence, the new business model ontology, by considering customer care strategy as the relationship mechanism, has 14 elements: Capabilities, Partnership, Value Proposition, Value Configuration, Customer, Relationship, Channel, Revenue, Cost, Information and Knowledge, Values and Emotions, Environment and Conditions, Thoughts and Perceptions, Preferences.

A business model ontology is a conceptual tool that includes a set of elements and their relationships, making it possible to express the logic of a specific company's business.

As previously mentioned, a business model consists of four main components. Using a cognitive approach in the customer relationship section can increase customer satisfaction and establish better and more sustainable relationships. The overall structure of the cognitive business model proposed in this paper is shown in [Fig. 4].

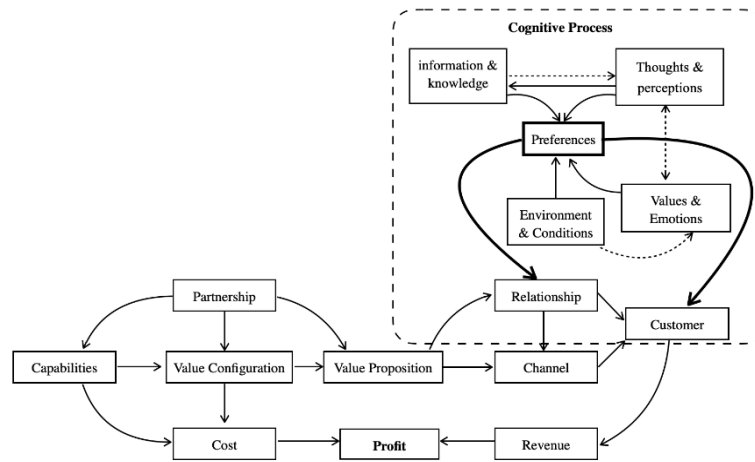

*Figure 4: A Cognitive Business Plan Ontology*

This design is based on the overall structure of the business model ontology [Currie 2004] and includes the following concepts:

*Value Proposition*: A value proposition provides an overview of the company's product and service packages that together create value for a specific customer segment.
- Represents the value for the target customer.
- Is based on capabilities.

This set includes one or more offers. A simple offer consists of part of the company's product and service set:
- Includes a description.
- Includes reasoning.
- Has a lifecycle.
- Has a value level.
- Has a price.

*Customer*: The target customer segment defines the types of customers that a company aims to serve.
- Receives the value proposition.

It consists of a set of criteria. A primary criterion defines the characteristics of a target customer group.

*Channel*: Describes how a company communicates with and reaches its customers. Its goal is to deliver the right amounts of the available products and services to the right people at the right place and time—considering cost, investment, and flexibility constraints.
- Delivers the value proposition.
- Reaches the target customer.

This set includes one or more channel links. A channel link describes part of a company's channel and shows specific marketing roles. Different channel links may sometimes communicate to allow for multi-channel cooperation.
- Inherits characteristics from the value proposition.
- Delivered by the executor.

*Relationship*: The relationship component explains the relationship a company establishes with the customer segment to create customer benefits through acquisition, retention, and sales growth.
- Considers the value proposition.
- Is based on the target customer.
- Participates in customer rights.

This set includes one or more mechanisms. A relationship mechanism describes the functional interaction between the company and its customers. It may also be a channel link or part of the value proposition.
Has a function.
- Inherits components from the channel link.

*Capability*: Describes the ability to execute a repeatable pattern of actions. The company must have several capabilities to deliver its value proposition. Capabilities are based on the company's or its partners' resources.
- Provides the value proposition.
- Consists of a set of resources.

Resources are inputs to the value creation process. They are the source of the company's needs to provide its value proposition.
- Are of various types.
- Delivered by the executor.

*Value Configuration*: Describes the arrangement of one or more activities to deliver a value proposition.
- Delivers the value proposition.
- Relies on capabilities.
- Has types.

The value structure consists of a set of activities. An activity is an action the company takes to perform work and achieve its goals.
- Has a level.
- Has a nature.
- Requires resources.
- Executed by the executor.

*Partnership*: A voluntary collaboration agreement formed between two or more independent companies to carry out a specific project or activity by coordinating necessary equipment, resources, and activities.
- Supports the value proposition.
- Relies on capabilities.

This set includes one or more agreements. An agreement specifies the performance and terms and conditions of participation with the executor.
- Includes reasoning.

- Has a degree of integration.
- Has a degree of dependency.
- Made by the executor.

*Information and Knowledge*: Information and knowledge influence preferences and have a reciprocal relationship with thoughts and perceptions.
- Affects preferences.
- Influenced by them.

*Emotions*: Values and emotions have various dimensions. They are not the same among different individuals, changing through learning and experience, and have a two-way relationship with thoughts and perceptions.

Environment and Conditions: Environment and conditions affect customer behaviour and decision-making directly and indirectly [through influence on preferences].

*Thoughts*: Thoughts and perceptions, while influenced by information, knowledge, values, and emotions, also have a direct impact on information, knowledge, and preferences.

*Preferences*: Preferences and decision-making are variable and influenced by various factors.
- Influenced by values and emotions.
- Influenced by environment and conditions.
- Influenced by information and knowledge.
- Influenced by thoughts and perceptions.

Preferences affect the target customer segment and customer relationship.

*Revenue*: Describes the method by which a company earns revenue and can consist of one or more elements of revenue streams and pricing.
- Part of financial aspects.

A *revenue* model is based on and related to the company's value proposition.
- A set of revenue streams and prices.
- Has characteristics of revenue stream and price.

*Cost*: *This* element measures all the monetary costs incurred by the company.
- Is a part of financial aspects.
- Consists of a set of accounts.

An account *records* the financial transactions (costs) of a specific category.

Profit: This *section* is the result of the business model configuration. The company's revenue model and cost structure together determine the company's profit or loss logic *and*, consequently, its ability to survive in competition.

The above *concepts* can be described mathematically and concisely using the statements in [Fig. 5 and Fig. 6], as follows:

| | | |
|---|---|---|
| VP rep(TC) base(CP) | VC rely(CAP) | PRF inf-of(INF) |
| VP base(CP) | PRT sup(VP) | PRF inf-of(EM) |
| TC recv(VP) | PRT rely(CAP) | PRF inf-of(EN) |
| TC comp(CR) | INF inf-on(PRF) | PRF inf-of(TH) |
| CH delv(VP) to (TC) | INF inf-of(PRF) | CS estab(VC) |
| TCR con(VP) | EM estab(TH) | CS estab(CAP) |
| TCR estab(TC) | EN inf-on(PRF) | RV rely(CQ) |
| TCR cont(CQ) | TH inf-on(PRF) | PF rely(RV) |
| CAP allo(VP) | TH inf-on(INF) | PF rely(CS) |
| VC prov(VP) | TH inf-of(INF) | |
| VP= Value Proposition | CAP= Capability | TH= Thoughts |
| TC= Target Customer | VC= Value Configuration | PRF= Preferences |
| CP= Capability | PRT= Partnership | RV= Revenue |
| CH= Channel | INF= Information and Knowledge | CS= Cost |
| TCR = Relationship | EM= Emotions | PF= Profit |
| CQ= Customer Segment | EN= Environment | |

*Figure 5: Meanings of terms used in the proposed ontology*

| | | |
|---|---|---|
| rep(_a): represents value for | conc(_obj): it concerns a | rely(_obj): it relies on |
| comp(_x): is composed of | estab(_obj): it is established with | has(_n): it has a type |
| base(_b): is based on | cont(_obj): it contributes to | sup(_obj): it supports the |
| recv(_obj): it receives a | Set-of(_x): it set of | inf-on(_obj): it influences on |
| delv(_obj): it delivers a | allo(_obj): it allows to provide the | inf-of(_obj): it influences of |
| delv(_obj_to): it delivers to a | prov(_obj): it provides | |

*Figure 6: Symbols and terms used in the proposed ontology*

## 5    Implementation of the Proposed Method using BEL model

To evaluate the proposed model, it is necessary to compare its performance with the original model. For this purpose, we utilized data provided by Samsung. The primary advantage of this dataset is its annual publication and public availability, which ensures a comprehensive and up-to-date collection of information. Moreover, the consistency and reliability of the data enhance the robustness of our findings, as it minimizes potential biases and inconsistencies that may arise from other data sources.

Therefore, data *related* to the new elements in the proposed ontology were implemented and evaluated using the BEL machine.

Initially, the original ontology model with 9 inputs [Capabilities, Partnership, Value Proposition, Value Configuration, Customer, Relationship, Channel, Revenue, Cost] and the prediction of Profit as the output was executed; then, the proposed ontology was run on the *machine*. In this stage, the numbers related to the 14 elements of the proposed cognitive business model ontology [Bagherzadeh, Setayeshi et al. 2023] (Capabilities, Partnership, Value Proposition, Value Configuration, Customer, Relationship, Channel, Revenue, Cost, Information and Knowledge, Values and Emotions, *Environment* and Conditions, Thoughts and Perceptions, Preferences) were used as the model inputs, and the value corresponding to Profit was considered as the system output.

Since the numbers associated with input elements [the business model ontology] each have different *scales* and sizes, it is necessary to preprocess the data first and bring them to a uniform scale. This is done in the Thalamus part. Then, as the data enter the Sensory Cortex section, feature selection and extraction are performed, extracting suitable and more effective features for entering the neural network. In the next stage, the prepared data enter the two neural networks present in the Orbitofrontal Cortex and Amygdala, and their outputs are integrated in the Amygdala. The final output is obtained by merging the data from these two sections.

Step 1: Data Preprocessing (Thalamus)

This step is crucial in the input section. To start, it is essential to transform the data into a manageable structure for analysis. This includes the following tasks: data cleaning and data normalization.

Step 2: Model Selection (Sensory Cortex)

Given the type of data and the prediction goal, different models can be used, each with its own advantages and disadvantages. Given the large volume of data and the complex relationships between them, neural networks were utilized.

Step 3: Model Implementation (Orbitofrontal Cortex and Amygdala)

In this step, the selected model is implemented. The data is initially divided into training and test sets to train the model and then test its accuracy.

## 6    Evaluation

After implementing the model, its performance must be evaluated. The evaluation criteria include:
- Mean Squared Error (MSE): To check the model's error rate.
- Coefficient of Determination ($R^2$): To assess the model's accuracy in prediction.

- Prediction vs. actual data charts: To visually examine the model's performance.

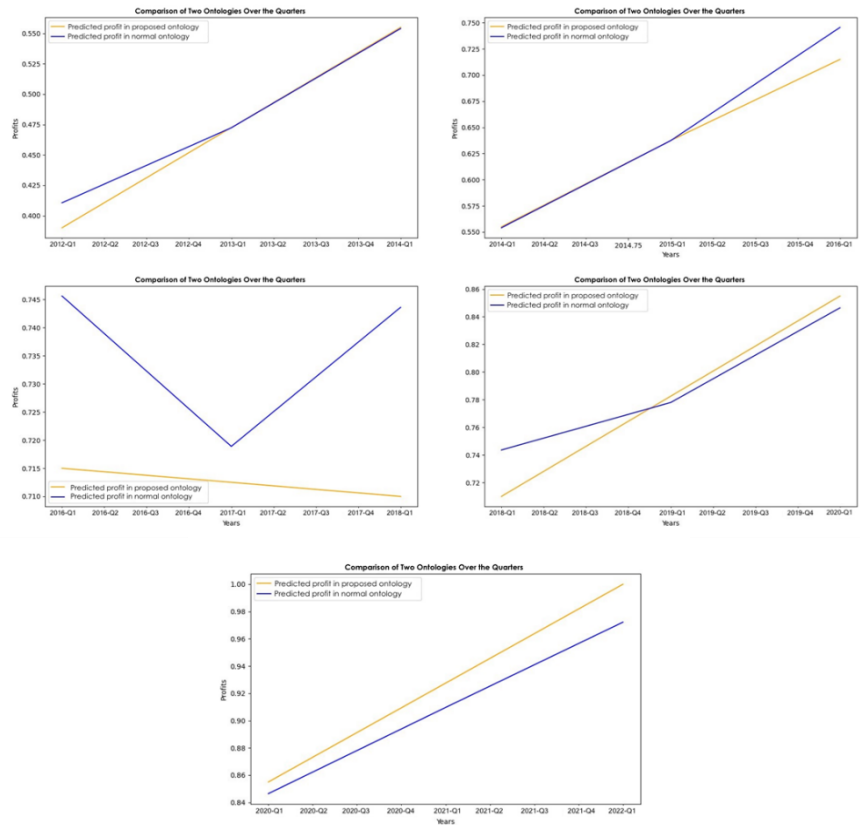

*Figure 7: Comparison of two Ontologies*

By incorporating these steps, we will analyse the results and provide the best model for profitability prediction.

Model Error and Accuracy in Predicting Profit Based on 14 Variables
- Polynomial Regression MSE: 4.879188223809868e-05
- Polynomial Regression R²: 0.9981379850923756

Model Error and Accuracy in Predicting Profit Based on 9 Variables
- Polynomial Regression MSE: 0.0004850535400444231
- Polynomial Regression R²: 0.9814891969497855

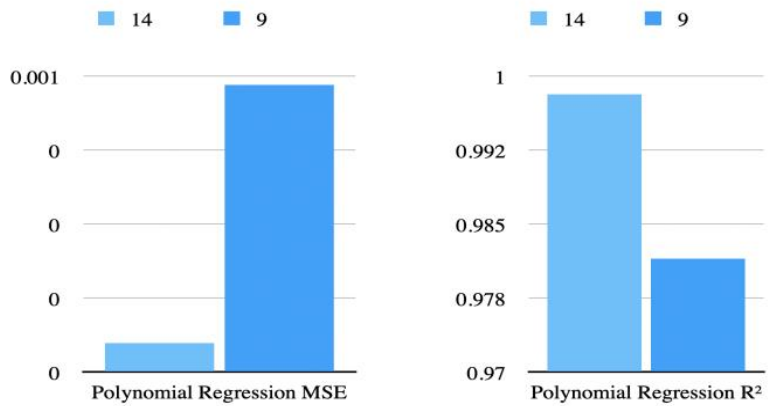

*Figure 8: Polynomial Regression of two models*

## 7 Results analysis

As it's observable in [Fig. 9] in the first year, 2013, the rate of change of actual profit reached its highest value while the rate of change of predicted profit remained steady. The rate of change of actual profit decreased significantly, dropping to 0.03 in 2014. During these years, the rate of change of actual profit showed considerable fluctuations, with notable drops in 2014 and 2016. Then, the rate of change of predicted profit remained steady at about 0.05 and the rate of change of actual profit became relatively stable, fluctuating between 0.02 and 0.03; the rate of change of predicted profit also showed a slight decline but remained around 0.025.

As observed, the rate of change in actual profit experienced significant fluctuations over the years. These fluctuations could be attributed to various economic factors, market conditions, and changes in supply and demand. Meanwhile, compared to the actual profit, the rate of change in predicted profit has been more stable and less prone to fluctuations. This stability could be due to the use of reliable prediction models and consistent data analysis methods.

This analysis highlights that actual profit is more susceptible to fluctuations, whereas predicted profit remains relatively stable. This underscores the importance of accurate analysis and prediction models to mitigate the risks associated with profit volatility.

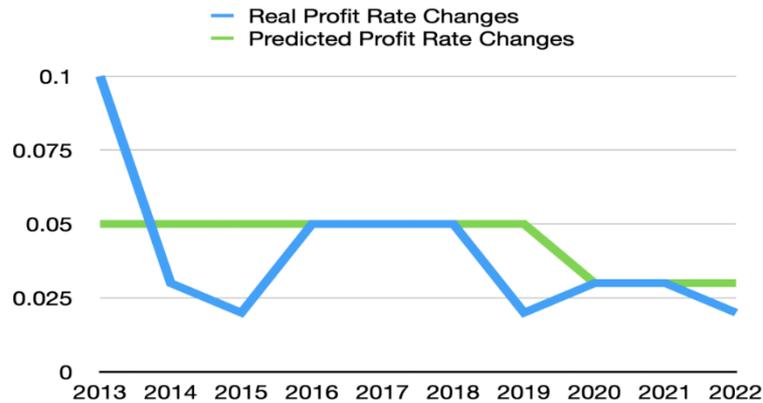

*Figure 9: Profit rate changes*

## 8 Conclusion

The use of ontology in any field can make a significant difference. One of the areas where the need for ontology is inevitable is the design of business models. Today's business landscape is accompanied by complexity and uncertainty, and the need for tools to deal with these issues is crucial for business advancement. Precise business planning for profitability, given the intense competition in the modern world, is essential. This means that all parts of the business must be optimized and strengthened. Given the necessity of designing ontology, it has also been addressed in the business roadmap discussion; optimizing and improving these ontologies can greatly assist in business management.

Cognitive sciences have garnered much attention from researchers today. Using cognitive sciences in the field of improving ontology can also yield more accurate results and better performance. Although business processes have been examined from a cognitive perspective at various stages, and solutions have been provided in these areas, this topic has been less focused on at the ontology level. The aim here is to define a general cognitive model for describing business models. Such an approach can have significant impacts in the future. The cognitive business model ontology to be presented, along with the tools built on it, represents the first step towards facilitating management under uncertainty.

Overall, the goal of this research is to address the concept of strategic business plans with a cognitive ontology approach. By focusing on the cognitive perspective and examining and identifying key concepts and relationships in the field of business plans based on cognitive sciences, and producing precise textual definitions and identifying conditions for such concepts and relationships, and by aligning these elements, a cognitive ontology for the business roadmap was presented.